\documentclass[aps,prl,reprint,superscriptaddress,nofootinbib]{revtex4-2}

\usepackage{amsmath,amssymb,bm,mathtools}
\usepackage{balance}
\usepackage{graphicx}
\usepackage{tikz}
\usetikzlibrary{arrows.meta,positioning,calc,decorations.pathmorphing}
\usepackage[colorlinks=true,citecolor=blue,urlcolor=blue,linkcolor=blue]{hyperref}

\usepackage[titletoc,toc,title]{appendix}

\newcommand{\dd}{\mathrm{d}}
\newcommand{\ii}{\mathrm{i}}
\newcommand{\cA}{\mathcal{A}}
\newcommand{\cF}{\mathcal{F}}
\newcommand{\cK}{\mathcal{K}}
\newcommand{\cR}{\mathcal{R}}
\newcommand{\cD}{\mathcal{D}}
\newcommand{\eps}{\epsilon}
\newcommand{\mpl}{\ell_{\rm P}}

\newcommand{\be}{\begin{equation}}
\newcommand{\ee}{\end{equation}}
\newcommand{\eq}[1]{(\ref{#1})}
\newcommand{\bit}{\begin{itemize}}  \newcommand{\eit}{\end{itemize}}
\newcommand{\ben}{\begin{enumerate}}  \newcommand{\een}{\end{enumerate}}

\newcommand{\ket}[1]{|#1 \rangle}

\def\bea{\begin{eqnarray}}
\def\eea{\end{eqnarray}}

\def\Tr{{\rm Tr}}
\def\one{\mbox{1 \kern-.59em {\rm l}}}

\def\a{\alpha}

\def\d{\delta}

\def\o{\omega}

\def\s{\sigma}  
\def\t{\tau}

\DeclareMathAlphabet\mathbfcal{OMS}{cmsy}{b}{n}

\def\cA{{\cal A}}  
\def\cD{{\cal D}}  \def\cF{{\cal F}}
  
\def\cK{{\cal K}} 
  
  \def\cR{{\cal R}}

\def\oh{\hat{\omega}}

\newcommand{\nn}{{\nonumber}}

\def\d{\delta}

\def\uno{\mbox{1 \kern-.59em {\rm l}}}

\def\one{1\!\!1\,\,}

\def\Box{\square}

\def\bcomment#1{}

 \def\ii{{\rm i}}

\def\IR{\relax{\rm I\kern-.18em R}}

\begin{document}

\title{From Horizon Microstates to the Black Hole Membrane}

\author{Chong-Sun Chu}
\affiliation{Department of Physics, National Tsing Hua University,
  Hsinchu 30013, Taiwan}
\affiliation{Center for Theory, Computation, and Data Science Research,
  National Tsing Hua University, Hsinchu 30013, Taiwan}
\affiliation{Physics Division, National Center for Theoretical Sciences,
  Taipei 10617, Taiwan}

\begin{abstract}

The membrane paradigm represents a black-hole horizon by a fictitious
conducting surface. We derive a microscopic electromagnetic membrane
from black-hole matrix quantum mechanics. The fuzzy-sphere horizon
carries a Berry monopole, placing its fundamental fermionic partons in
lowest-Landau-level states with Ohmic and Hall responses. Off-diagonal
bifundamental modes connecting the horizon and exterior matrix blocks
become tachyonic near the horizon and condense, dynamically coupling
the horizon gauge field to the exterior Maxwell field. In the low-frequency
regime
$\omega R \ll1$, the fixed-parton transport description predicts
frequency- and helicity-dependent reflectivity. At larger frequency,
real parton excitations require a black-hole $S$-matrix. Ohm's law then
fixes the inclusive absorption probability; unitarity bounds the local
absorption cross section by the horizon area, with the classical
conductivity $1/4\pi$ saturating this maximal-absorption bound.

\end{abstract}

\maketitle
\paragraph*{Introduction.---}

The black-hole membrane paradigm encodes the response seen by an
exterior observer in terms of a timelike stretched horizon endowed
with surface charge, current, and transport coefficients
\cite{Damour:1978cg,MacDonald:1982zz,
  Price:1986yy,Thorne:1986iy}. For
Maxwell theory the membrane current, derived as a boundary source for the
bulk Maxwell field \cite{Parikh:1997ma}, satisfies
\be
  j^{\rm mem}_{A}=\frac{1}{4\pi}F_{A\mu}n^\mu,
  \label{classical_surface_current}
  \ee
where $n^\mu$ is the outward normal.
The constitutive  Ohm's law
$j^{\rm mem}_{A}= E_{A}/4\pi$, written in the FIDO (Fiducial Observer) frame,
is equivalent to the purely ingoing
horizon boundary condition.
In the classical construction, however, $j^{\rm mem}_{A}$ is not an
independent microscopic current, but is introduced to reproduce the
dynamics seen by an exterior observer. 
A quantum membrane paradigm requires two ingredients that the
classical construction does not provide: physical horizon states
carrying a current, and a dynamical mechanism that makes this current
source the exterior field.
Here we derive both ingredients in the  $SU(N)$ matrix quantum mechanics
proposed in Ref.~\cite{Chu:2024qil}.

The model contains
three $N\times N$ traceless matrix Hermitian coordinates $X_a$, $a=1,2,3$, 
and two-component $N \times N$ matrix fermions
$\psi$.  $X_a$ transforms in the adjoint, while
$\psi$ transforms in the fundamental representation of
$SU(N)$ and so
the second matrix index of $\psi$ is a flavor index rather than a second index
of an adjoint representation.
The action  is 
\begin{align} \label{model}
 S={}& \int dt\,  \Tr \Bigl[\frac{\dot X_a^2}{2a_0^2M_{\rm P}}
 +\frac{M_{\rm P}}{N^2}\bigl([X_a,X_b]^2+4X_a^2\bigr)\nonumber\\
 &\hspace{1.7cm}+\ii\psi^\dagger\dot\psi
 -\frac{a_2M_{\rm P}}{N^2}\psi^\dagger\sigma_aX_a\psi\Bigr],
\end{align}
where $M_{\rm P}$ is a mass scale (Planck) of the theory, $a_0,a_2$ are
parameters that are to be fixed by requiring the theory to reproduce
the expected properties of black holes.
Unlike the adjoint-matrix constructions of BFSS and
IKKT~\cite{Banks:1996vh, Ishibashi:1996xs},
the negative mass term supports a rank-$N$
fuzzy-sphere solution $X_a=J_a$, with
\be
 [J_a,J_b]= i\eps_{abc}J_c,
 \qquad J=\frac{N-1}{2},
 \label{fuzzy}
\ee
It was found that a
fuzzy sphere together with a half-filled Fermi sea reproduces the
horizon radius $R= N\mpl$
of a Schwarzschild black hole, as well as the Bekenstein-Hawking
entropy of black hole as a counting of the horizon microstates in the Fermi sea
\cite{Chu:2024qil}.
The same framework has reproduced the size, shape, entropy and the
angular momentum of the quantum Kerr black hole \cite{Chu:2024edh}, as well
as  the Hawking decay
rate of a semi-classical black hole \cite{Chu:2026wyh}
as a nonperturbative tunneling of fuzzy spheres in the matrix model
\cite{Chu:2026wyh}. It is nontrivial that the same condition $a_0= \pi/3$ was
required in the matching of the angular momentum of Kerr
black hole, and, independently, the Hawking decay rate of Schwarzschild
black hole.
Classically the fluctuation spectrum over the fuzzy sphere contains a tachyonic
mode \cite{Chu:2024qil}. This is however stabilized fully by 1-loop quantum
effects \cite{chaos}.

\begin{figure}[t]
  \centering
  \includegraphics[scale=0.25]{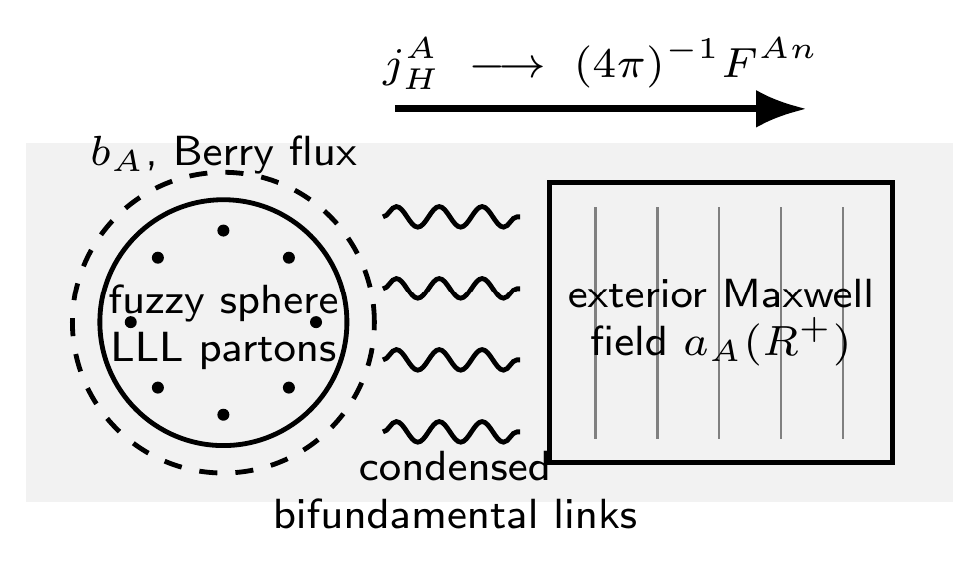}
  \caption{Microscopic origin of the membrane current. Fundamental
    horizon partons form LLL states on the fuzzy sphere and carry an
    Ohmic and Hall current.  A condensate of off-diagonal links
    occupies a microscopic layer of coordinate thickness $\d r
    \lesssim \mpl$ and dynamically couples the horizon gauge field
    $b_A$ to the boundary value of the exterior field $a_A$.  The
    condensate thereby makes the total horizon current the boundary
    source of the exterior Maxwell field.}
\label{mechanism}
\end{figure}

In this letter, we show that the same spin-$J$ structure that
quantizes the sphere also produces a Berry monopole, and that the
fundamental fermions form lowest-Landau-level (LLL) horizon
partons. We then place the horizon and the exterior Maxwell region in
two diagonal matrix blocks. The off-diagonal link fields condense near
the horizon and couple the two gauge systems, allowing the microscopic
membrane current to be accessible to the external
observer, thereby replacing the classical fictitious membrane with a
quantum membrane located in the vicinity of the horizon.
The microscopic mechanism is  summarized in Fig.~\ref{mechanism}.

\paragraph*{Lowest-Landau-level horizon partons.---}
Let $|\Omega\rangle$ be a spin-$J$ coherent state of the fuzzy sphere. The
intrinsic phase ambiguity in its definition gives rise to the Berry connection
\be
  \cA =-\ii\langle\Omega|\dd|\Omega\rangle
  =J(1-\cos\theta)\dd\phi,
  \ee
  whose first Chern number is
  \be
  \frac{1}{2\pi}\int_{S^2}\dd\cA=2J=N-1.
  \label{berry}
\ee
i.e. it is a monopole of charge $J$. This connection cancels between
the two indices of an
adjoint matrix, but not for a fundamental fermion.  The
fermions of Ref.~\cite{Chu:2024qil}, however, couple to the fuzzy sphere
by left multiplication and therefore transform as $N$ flavors of
fundamental spin-$J$ states.  More explicitly, the coherent-state
resolution of the identity is $\mathbf
1_N=(N/4\pi)\int\dd\Omega\,|\Omega\rangle\langle\Omega|$. On a sphere
of radius $R$, the continuum field $\chi_{\alpha I}=\sqrt{N/(4\pi
  R^2)}\langle\Omega|\psi_{\alpha I}\rangle$ has a canonically
normalized first-order kinetic term. Its orbital wave functions are
$P(z)/(1+|z|^2)^J$, with $\deg P\leq 2J$, and therefore span precisely
the finite monopole
Hilbert space.  The finite spatial Hilbert space is an interesting feature
of the model: it is not a regulator
imposed on a continuum fermion; it follows directly from the matrix
representation.

Combining the orbital spin $J$ with the Pauli spin gives total spins
$K_u=J+1/2$ and $K_v=J-1/2$. After particle-hole conjugation of the
lower band, the fermion action of \eq{model} gives
two positive-energy parton sectors $u$ and $d$, which,
due to the presence of the instrinsic Berry monopole, becomes LLL
sections  with Chern numbers
\be
 k_u=N, \qquad k_d=-(N-2) \simeq -N,
 \label{chernud}
\ee
and degeneracies $N+1$ and $N-1$ for every flavor. Their spatial wave
functions obey $D_{\bar z}^{(k_s)}s_I=0$ rather than an ordinary
second-order wave equation. Including the $N$ flavor copies gives
$N(N+1)$ upper-band states and $N(N-1)$ lower-band states, or $N^2$
states each in the leading order of large $N$.
A quantum black-hole is identified with the half-filled ensemble
$r+s=N^2$ of horizon partons with  $r$ u-partons and $s$ d-partons.
The model has a conserved $U(1)$ symmetry $\psi \to e^{i \a} \psi$ which implies
that  the electric charge $Q:=q(r-s)$ is conserved. Thus the same states
that account for the black hole microscopic
entropy also carry the charge entering the transport response. We note that
the action of the horizon partons contains the 
occupation-energy scale
\be \label{parton_gap}
 \epsilon_0=\frac{a_2}{4a_0\mpl}=O(\mpl^{-1})
 \ee
 which control the  creation of partons and their virtual process.

To facilitate the discussion of the Ohm's law, we introduce the FIDO proper time
$\t$ which is related to the time $t$ of the model \eq{model}
as
\be
\t = \frac{2 a_0 M_P \mpl}{N}t.
\ee
Here the $1/N$ redshift factor  is  
derived from
the $g_{00}$-component of the Schwarzschild metric at the location of the
  stretched horizon.
For an individual parton represented by a
coherent packet centered at the location $\xi^A(\t)$ on the fuzzy sphere, 
the fermionic action of the model gives the first-order action for its
guiding-center
\be
  S_{\rm LLL}=\pm\int\dd\tau\,
  \bigl(\cA_A\dot\xi^A-qR b_A\dot\xi^A\bigr),
  \label{LLL_action}
\ee
where $q := a_2/(2a_0)$ denotes the charge of the partons,
$b_A$ is the horizon gauge field and the $\pm$
sign is for the $u/d$ partons. The opposite first-order structures
parallel the two oppositely charged endpoints of an open string in a
background two-form field~\cite{Chu:1998qz}. An applied tangential
electric field therefore causes Hall drift. Since the
black-hole horizon is conjectured to be a thermal system
with a Hawking temperature $T_H = 1/4 \pi R$,
the remaining horizon degrees of freedom act as an effective bath for each
parton.
As a result, we propose to describe the dynamics
of the partons with the first-order
Langevin equation of the form
\be
\cF_s\eps_{AB}v_s^B=-q_sE_A+\eta v_{sA},
 \qquad \cF_s=\pm\frac{J}{R^2},\quad q_s=\pm q,
 \label{langevin}
\ee
where the correlated signs apply to the $u/d$-sectors and
$\eta$ is a friction coefficient arising from the thermal relaxation.
A reasonable estimate is $\eta\sim T_H^2\sim N^{-2}\mpl^{-2}$.
Solving Eq.~\eqref{langevin} and summing the occupied states gives
the linear response
\be  \label{constitutive}
  j^A=
  \sigma_{xx} E^A  +\sigma_H\eps^{AB} E_{B},
\ee
where $E^A=-\partial_\tau b^A$ is the electric field on the horizon.
Since $\cF_s\sim N^{-1}\mpl^{-2}$, this 
gives $\eta/\cF_s\sim N^{-1}$ and the longitudinal conductivity
$\s_{xx} \simeq Nq^2 \eta/ (2 \pi \cF_s) = O(N^0)$, whereas
\be
  \sigma_H=\frac{qQ}{2\pi N}+O(N^{-2})
  \label{hall}
\ee
for black-hole of charge $Q$.

The transport law \eqref{constitutive} assumes that the applied field
does not create real horizon partons. The fixed-parton-number,
one-body guiding-center description is therefore controlled for
$\hat\omega\ll\epsilon_0$, or
\be \label{low}
 \oh \mpl\ll 1.
 \ee
 Here $\hat\omega$ is the local FIDO frequency of the applied field,
 $b_A \sim e^{-i \oh \t}$. Since the microscopic $U(1)$ charge is
conserved, the first real excitation is a particle--hole pair with
threshold $\Delta_{\rm pair}=\epsilon_u+\epsilon_d\simeq2\epsilon_0$.
Once $\hat\omega$ approaches this threshold, sectors with different
parton occupations become accessible and the fixed-parton-number
constitutive description is no longer complete.

Equation \eqref{constitutive} is a genuine current of microscopic
degrees of freedom, but it initially couples to a gauge field $b_A$
living on the fuzzy sphere. The exterior observer instead measures a
Maxwell field $a_A(r,\Omega)$. A membrane paradigm therefore requires
a dynamical relation between these fields.

\paragraph*{A dynamical horizon--environment interface.---}
To couple the intrinsic response to an exterior observer, consider a
two-block matrix configuration \cite{Chu:2026vzj}
\be
  \bm X_a=
  \begin{pmatrix}
    X_a^{H} & W_a\\
    W_a^\dagger & X_a^{E}
  \end{pmatrix},
  \label{two_block}
  \ee
  where the  black-hole block $X_a^H$ is of rank $N$ and describes the
  physics of the fuzzy sphere, the environmental
  block $X_a^E$ is of rank $N' \to \infty$ and it describes a Maxwell
  theory outside the fuzzy sphere. The two blocks interact through
  the off-diagonal link block $W_a$, whose
  Lagrangian takes the form
\be\label{L-link}
L_W \sim
  \Tr_H \dot W_a \dot W_a^\dagger -U_2 -U_4.
\ee
Here $U_2$ is quadratic in $W$, and  includes the
  mass term as well as the
  linear and quadratic interaction of $W$ with the
  gauge fields;
  and $U_4$  denotes the quartic self-interaction of the $W$'s. When the quartic
  potential is positive, 
  a condensate  for the link field can occur if
  the quadratic term has negative mass. The condensate is most effective
  in the direction where the mass term is most negative.
  A simple analysis of the quadratic link operator reveals that
  at a fixed radial direction, the most unstable polarization is
  one-dimensional.  This
  eigenmode
  is tangential to the fuzzy sphere and has the maximal helicity $ J_3 =J$ and
  $S_3 = 1$ with respect to the angular momentum and coordinate rotation.
  The  corresponding eigenvalue is
  \be
  k_-(\xi)=2(\xi_a-J_a)^2-4(J+1),
  \label{eigenvalue}
\ee
where $\xi_a = \frac{x_a}{2 \mpl} + L p_a$ and $x_a, p_a$ labels the position and
momentum of the environmental state.
Here $L= N' \mpl/N$ is an IR regulator length that characterizes the size of
the environmental space outside the black hole.
The quartic potential in this channel is positive and we have
\be
  U(w) \propto \bigl[k_-(\xi)|w|^2+2|w|^4\bigr].
  \label{linkpotential}
\ee
Therefore the tachyonic mode condenses whenever $k_-<0$.
At a fixed radial direction,
the most effective minimum occurs at $\xi_a=J n_a$.
Translating this condition into the emergent radial coordinate shows that
the condensate occupies a microscopic
layer of thickness $\d r = O(\mpl)$ immediately outside
the horizon. This gives a microscopic identification of
the stretched horizon.

The rotationally populated condensate produces the gauge-invariant interface
potential term
\be
  V_{\rm lock}= \frac{\cK}{2}
  \int \dd\Omega\,
  (b_A-a_A +\partial_A\vartheta)^2,
  \qquad
  \cK=\frac{N^3\mpl}{8\pi a_0},
  \label{lock}
  \ee
  where $\vartheta$ is the phase of the link condensate.
  In unitary gauge $\vartheta=0$.
The term penalizes the relative gauge field rather than either gauge
field separately. An infinite stiffness $\cK \to \infty$
implies that the two gauge fields are
perfectly locked. In general,
kinetic energy may shake the configuration out of its minimum.
For a finite but large $\cK$, locking $b_A\simeq a_A(R^+)$ is effective for
slowly varying fields. 

\paragraph*{Transfer of the microscopic parton current.---}
The physical material current in the matrix model is the current
\eq{constitutive} carried
by the LLL partons,
\be \label{parton_current}
 j_A[b]=\sigma_{xx}E_A+\sigma_H\eps_{AB}E^B,
 \qquad E_A=-\partial_\tau b_A .
\ee
The horizon gauge field also has its own quadratic action. The action for the
partons-gauge system is
\begin{align}
 S={}&-\frac{1}{16\pi}\int_{\rm out}\dd^4x\sqrt{-g}\,
 F_{\mu\nu}F^{\mu\nu}+S_H[b]+S_{\rm parton}[\psi,b]\nonumber\\
 &-\frac{\cK}{2}\int\dd\tau\dd\Omega\,(b_A-a_A)^2,
 \label{effective_action}
\end{align}
where
\be
 S_H[b]=\frac{R^2}{2g_{\rm BH}^2}\int\dd\tau\dd\Omega
 \left[(\partial_\tau b_A)^2-\frac{1}{\mpl^2}f^2\right]
\ee
is the action for the emergent gauge field on the fuzzy sphere. Here
$f$ is the magnetic field and 
the gauge coupling ${1}/{2 g_{\rm BH}^2} = {\mpl}/{4\pi a_0}$.
Varying $a_A$ and $b_A$ gives
\begin{align}
  &\frac{1}{4\pi}F_{A n}+\frac{\cK}{R^2}(b_A-a_A)=0,
  \label{var_a}\\
  &\frac{1}{g_{\rm BH}^2}
  (\Box b_A-\partial_A\partial_B b^B)
  -j_{A}-\frac{\cK}{R^2}(b_A-a_A)=0,
  \label{var_b}
\end{align}
where the parton current
\be
j_A := -\frac{1}{\sqrt{-h}}\frac{\d S_{\rm parton}}{\d b^A}
\ee
is given by Eq.~\eqref{parton_current}.
The same relative gauge field term appears with opposite signs.
Eliminating it yields the stiffness-independent relation
\be \label{central}
\frac{1}{4\pi}F_{An}= j_A^{\rm mem},
\ee
where
\bea \label{c1}
&& j^{\rm mem}_{A} = j_{A} + j^{\rm pol}_{A}, \nn\\
&& j_A^{\rm pol} :=  -\frac{1}{g_{\rm BH}^2}
 \left(\Box \d_{AB}-\partial_A\partial_B\right)b^B 
\eea
is a polarization current that arises from the kinetic motion
  of the gauge field.
Thus the microscopic horizon response is not merely analogous to the
membrane current: it is transferred dynamically as the physical
boundary source of the exterior Maxwell field. We emphasize that
condensation is
essential, but the transfer relation \eqref{central} does not require
an infinite
$\cK$. We note also that taking the tangential divergence of
Eq.~\eqref{central} gives the boundary continuity equation inherited
from exterior Maxwell gauge invariance; equivalently, the link
condensate transfers charge without violating the full matrix-model
Gauss constraint.

In the regime where
fields are slowly varying with low energy and low angular momentum,
the locking becomes effective $b_A = a_A$ at the stretched horizon and
for low frequency \eq{low}
such that the fixed-number-number constitutive picture holds,
\eqref{central} becomes a boundary condition for the
{\it external} field. For an external field $a_A \sim e^{-i\oh \t} = e^{- i \o t}$,
where
\be
\o= \frac{2 a_0 M_P \mpl}{N} \oh 
\ee
is the frequency measured by an asymptotic observer using the time $t$
and $\oh$ is
the FIDO frequency,
the membrane current reads
\be\label{membrane_bc}
j^{\rm mem}_{A} = \sigma_{AB}(\o)E^B,
\ee
where
\bea \label{cc1}
& \sigma_{AB} =(\sigma_{xx}+ i \o D)\delta_{AB} +\sigma_H\eps_{AB} \nn\\[2mm]
& D = 
\frac{N}{4\pi a_0^2 M_P}, \quad \s_H = \frac{q Q}{2\pi N}. 
\eea
Here the reactive coefficient $D \sim R$ is generated by the horizon gauge 
kinetic term, the Hall term $\s_H$
arises from the LLL transport of the
partons induced by the intrinsic Berry monopole on the fuzzy sphere, and
the Ohmic term $\s_{xx}$ originates from the thermal dissipative effects
of the horizon. The classical neutral membrane is recovered when
$\sigma_{xx}=1/(4\pi)$ and $D=\sigma_H=0$. The $N$-independent scaling
of $\sigma_{xx}$ is robust and follows from the horizon relaxation scale.
Determining its exact coefficient requires a microscopic Kubo
calculation.
For a Schwarzschild exterior, the near horizon condition
\eqref{membrane_bc} gives the boundary condition
\be
  \partial_{r_*}a_A-4\pi\sigma_{AB}\partial_t a^B=0,
  \label{schw_bc}
  \ee
  where $r_*$ is the tortoise coordinate defined by $dr_* = dr/(1-R/r)$.
  In terms of circular polarizations  $\eps^A{}_B e_{(\pm)}^B=\pm\ii e_{(\pm)}^A$,
  the two admittances are
\be
  \sigma_\pm=\sigma_{xx}+\ii(\omega D\pm\sigma_H).
  \label{helicity_sigma}
  \ee
For each circular polarization the membrane boundary condition gives
the elastic reflection amplitude
\be
 \cR_\pm(\omega)=
 \frac{1-4\pi\sigma_\pm}{1+4\pi\sigma_\pm}.
\ee
Taking the classical Ohmic part as the absorbing reference, the
reactive terms generate the horizon reflection amplitude
\be
  \cR_\pm(\omega)=
  \frac{-\ii\cD_\pm}{1+\ii\cD_\pm},
  \qquad
  \cD_\pm=2\pi(\omega D\pm\sigma_H).
  \label{reflection}
  \ee
The reactive term makes a neutral horizon frequency dependent, while
the Hall term splits the two helicities and generates polarization
rotation and ellipticity. Unlike phenomenological models of
reflective horizon \cite{Berti:2025hly,
Cardoso:2016rao,Abedi:2016hgu,
Cardoso:2017cqb, Bambi:2015kza}, 
the electromagnetic
response of the horizon is fixed by
the transport coefficients of identified microscopic horizon states.

\paragraph*{Black hole S-matrix.---}

When $\hat\omega$ approaches the particle--hole threshold
$\Delta_{\rm pair}$, equivalently when $\omega D=O(1)$,
real parton excitations open the
channels
\be
\gamma(\o)+{\rm BH}_i\longrightarrow
\begin{cases}
\gamma_{\rm ref}(\o)+{\rm BH}_i,
& \text{elastic reflection},\\[1mm]
{\rm BH}_f+X_f,
& \text{inelastic absorption}
\end{cases}
\label{S_channels}
\ee
and the fixed-parton number constitutive description is no longer a complete one.
Nevertheless, at sufficiently weak
applied field, a frequency-dependent linear response function may still be
defined
\be
\langle J_A(\omega)\rangle = \s_{AB}(\o) E_B,\qquad
\sigma_{AB}(\omega)
=\frac{\Pi^R_{AB}(\omega)}{i(\omega+i0)}.
\ee
Here the retarded polarization tensor $\Pi^R_{AB}(\omega)$
is given by the Kubo formula,
namely the Fourier transform of the retarded current-current commutator.
This generalizes the low-frequency constitutive law \eq{membrane_bc}, \eq{cc1};
the membrane boundary condition
retains the same form, with $\sigma_{AB}$ replaced by the full
frequency-dependent response kernel $\sigma_{AB}(\omega)$.
Note that this is an inclusive description and it
does not specify which black-hole state was produced. The full on-shell
microscopic scattering information is encoded in the matrix-model
black hole S-matrix.
Let $S_{fi}$ denote the inelastic scattering amplitudes. Unitarity gives
\be \label{local_unitarity}
 |\cR|^2+\sum_f|S_{fi}|^2=1
 \ee
 and this gives a constraint on the transport coefficient $\s_{AB}$.
To see this, let us diagonalize $\s_{AB}$,
the membrane boundary condition \eq{central} then gives
\be \label{cR}
\cR(\omega)= \frac{1-4\pi\sigma(\omega)}{1+4\pi\sigma(\omega)}
\ee
for each eigen-channel. Now consider
a locally uniform incident wave and define the absorption cross section
$\sigma_{\rm abs}^{\rm loc}=A_H\cA$, where $A_H=4\pi R^2$ is the horizon area
and $\cA(\omega):=\sum_f|S_{fi}|^2
=1-|\cR|^2$ is the total absorption probability. This gives
\be
\cA(\omega) =\frac{16\pi\,\operatorname{Re}\sigma}
 {|1+4\pi\sigma|^2}.
 \label{inclusive_absorption}
 \ee
 Passivity,
${\rm Re}\sigma\geq0$, then implies the unitarity bound
\be
0\leq\sigma_{\rm abs}^{\rm loc}\leq A_H,
\ee
and
\be
 \sigma_{\rm abs}^{\rm loc}=A_H
 \ \Longleftrightarrow\ 
 \sigma=\frac{1}{4\pi}.
 \label{absorption_bound}
 \ee
 The classical black hole is therefore the
impedance-matched point that saturates the
unitarity bound. At larger frequency the full matrix-model black-hole
$S$-matrix is required to resolve the individual excited final states.

\paragraph*{Discussion.---}

We have derived a microscopic origin of the electromagnetic membrane
paradigm in matrix quantum mechanics. The topology of the fuzzy
horizon converts fundamental fermions into LLL partons with a real
surface current.
The bifundamental vector link sector produces a dynamical
interface; and the interface transfers the current to the exterior
Maxwell boundary condition.
The mechanism does not assume a
pre-existing material surface at the horizon. Instead, the stretched
horizon emerges as a microscopic region in which bifundamental
degrees of freedom condense. This provides a concrete sense in which
black-hole microstates can become operationally visible to an exterior
observer while recovering the form of the classical membrane response
at long wavelengths.

The derivation separates robust statements from dynamical quantities
still to be determined. The Berry topology, LLL Hilbert spaces,
tachyonic direction, positive single-mode quartic term, and
finite-stiffness current transfer are microscopic results. The
low-frequency regime $\omega D\ll1$ admits a
one-body guiding-center   transport
description and predicts frequency- and helicity-dependent reflection.
When the local frequency reaches the parton threshold, the same matrix
quantum mechanics instead supplies a multi-channel black-hole
$S$-matrix. Conductivity survives as an inclusive constraint on this
$S$-matrix, and the classical membrane is singled out as the
impedance-matched point of maximal local absorption.
A microscopic Kubo computation of $\sigma_{xx}(\omega)$
and a complete minimization over the unstable link subspace
are the principal quantitative next steps. The
black-hole $S$-matrix provides a natural framework for probing
the microscopic dynamics, inelastic channels, and unitarity of quantum
black holes.


\balance
\begin{acknowledgments}
We thank Sumit Das, Dimitrios Giataganas, Pei-Ming Ho, Norihiro
Iizuka, Satoshi Iso, Asuka Ito, Hikaru Kawai, Hsiang-Nan Li, Ian Low,
Emil Martinec, Samir Mathur, Jiro Soda, and Harold Steinacker for
discussions and comments. This work was supported by NCTS, the
National Science and Technology Council of Taiwan under Grant
No.~113-2112-M-007-039-MY3, and the National Tsing Hua University 2025
Talent Development Fund through a Tsai Wang, Yuan-Yang Distinguished
Talent Chair Professorship.

\end{acknowledgments}

\onecolumngrid
\appendix
\vskip 0.3cm
\centerline{-----------------------}
\section{A. Berry monopole and finite LLL Hilbert space}

Let $|J,m\rangle$ be the spin-$J$ irreducible representation with
$J=(N-1)/2$. An $SU(2)$ coherent states is defined by
\be
\ket{\Omega} := g(\Omega) \ket{J,J},
\ee
where $\Omega = (\theta, \phi)$ represents a point on $S^2$,
$g(\Omega) \in SU(2)$ is a rotation taking the north pole to $\Omega$.
It has the  Berry connection and curvature 
\be
 \cA=-\ii\langle\Omega|\dd|\Omega\rangle
   =J(1-\cos\theta)\dd\phi,
 \qquad
 \cF=\dd\cA=J\sin\theta\dd\theta\wedge\dd\phi,
\ee
and 
\be
 c_1=\frac{1}{2\pi}\int_{S^2}\cF=2J=N-1.
\ee
For a fundamental state $|\psi\rangle=\sum_m\psi_m|J,m\rangle$,
the coherent-state wave function 
\be
 \psi(z,\bar z)=\langle\Omega|\psi\rangle
 =\frac{P_{N-1}(z)}{(1+|z|^2)^J},\qquad
\mbox{where $P_{N-1}(z)$ is a polynomial of degree  $\leq N-1$},
\ee
obeys the covariant holomorphic equation
$D_{\bar z}\psi=0$ and is therefore a monopole LLL state. For an
adjoint matrix, the coherent-state phases of its two indices cancel;
this is why the Berry monopole couples to the fundamental partons but
not to ordinary adjoint fuzzy-sphere fluctuations.

Combining orbital spin $J$ with the Pauli spin gives total spins
$K_u=J+1/2$ and $K_v=J-1/2$. The operator $\bm\sigma\cdot\bm J$ has
eigenvalues
\be
 \bm \sigma\cdot \bm J=
 \begin{cases}
 J, & K=J+\frac12,\\ -(J+1), & K=J-\frac12.
 \end{cases}
 \label{sigmaJ_eigenvalues}
\ee
After particle-hole conjugation of the lower band, we obtain  two
positive-energy sectors: the $u$ and $d$ partons used in the
Letter.

\section{B. Guiding-center dynamics and current response}

For a localized parton packet the first-order action is
\be
 S_s=s\int\dd\tau\left(\cA_A\dot\xi^A-qR b_A\dot\xi^A\right),
 \qquad s=+1\;(u),\ -1\;(d).
\ee
The monopole field in an orthonormal frame is $\cF=J/R^2$.
Adding a phenomenological friction $\eta$-term gives
\be
 s \cF\eps_{AB}v_s^B=-s q E_A+\eta v_{sA}.
 \ee
Solving and multiplying by the charge density yields
\bea
 \sigma_{xx}^{(s)}&=&\frac{Nq^2\nu_s}{2\pi}
 \frac{\gamma_T}{1+\gamma_T^2},\\
 \sigma_H^{(s)}&=&s\frac{Nq^2\nu_s}{2\pi}
 \frac{1}{1+\gamma_T^2},
 \qquad \gamma_T=\frac{\eta}{\cF}.
\eea
The longitudinal terms add and the Hall terms subtract. With half filling,
$\nu_u+\nu_d=1$, and black-hole charge $Q=qN^2(\nu_u-\nu_d)$,
\begin{align}
 \sigma_{xx}&=\frac{Nq^2}{2\pi}
 \frac{\gamma_T}{1+\gamma_T^2},\\
 \sigma_H&=\frac{qQ}{2\pi N}
 \frac{1}{1+\gamma_T^2}.
\end{align}
The thermal scaling $\eta\sim T_H^2\sim R^{-2}$ gives $\gamma_T\sim
N^{-1}$ and hence $\sigma_{xx}=O(N^0)$. The exact computation of $\s_{xx}$
requires
a microscopic Kubo computation.

Note that the first-order transport law presumes that the applied field moves
the guiding centers without changing the occupied parton sector. This
restriction follows directly from the fermion spectrum. In fact rewriting the
fermion Hamiltonian in FIDO time,
\be
 \tau=\alpha_{\rm sh}t,
 \qquad
 \alpha_{\rm sh}:=\frac{2a_0M_P\mpl}{N},
 \label{redshift_factor}
\ee
we obtain
\be
 h_F=\frac{a_2}{2a_0N\mpl}\,\bm\sigma\cdot\bm J.
 \label{fido_fermion_hamiltonian}
\ee
Using Eq.~\eqref{sigmaJ_eigenvalues}, we find
the positive particle and hole energies
\be
 \eps_u=\frac{a_2J}{2a_0N\mpl},
 \qquad
 \eps_d=\frac{a_2(J+1)}{2a_0N\mpl},
 \qquad
 \eps_{u,d}=\eps_0+O(N^{-1}),
 \label{parton_energies}
\ee
where
\be
 \eps_0:=\frac{a_2}{4a_0\mpl}=O(\mpl^{-1}).
 \label{epsilon0}
\ee
Because the microscopic $U(1)$ charge is conserved, the simplest process
that changes the positive-energy parton content is a particle-hole excitation.
The threshold is parametrically given by
\be
 \Delta_{\rm pair}=\eps_u+\eps_d=2\eps_0+O(N^{-1}).
 \label{pair_threshold}
\ee
Consequently, the condition for a fixed-parton,
one-body guiding-center description is
\be
 \hat\omega\ll\eps_0\sim\mpl^{-1},
 \label{fixed_parton_condition}
\ee
where $\hat\omega$ is the local FIDO frequency. When this condition
is satisfied,  transitions between sectors with different
positive-energy parton occupation are suppressed. 

The asymptotic frequency $\omega$ is related to the FIDO frequency by
\be
 \omega=\alpha_{\rm sh}\hat\omega.
 \label{frequency_redshift}
\ee
The local reactive coefficient generated by the horizon gauge kinetic term is
\be
 \hat D=\frac{1}{g_{{\rm BH}}^2}=\frac{\mpl}{2\pi a_0},
 \label{D_local}
\ee
where $1/(2g_{{\rm BH}}^2)=\mpl/(4\pi a_0)$. The corresponding
asymptotic reactive coefficient  defined by the relation
$\omega D=\hat\omega\hat D$ gives
\be
 D=\frac{\hat D}{\alpha_{\rm sh}}
 =\frac{N}{4\pi a_0^2 M_P},
  \label{D_asymptotic}
\ee
Thus, for order-one microscopic couplings, the fixed-parton regime
gives
parametrically the low-frequency regime $\omega D\ll O(1)$ used in the Letter.

\section{C. Two-block action and link instability}

Write the matrix coordinate as
\be
 \bm X_a=\begin{pmatrix}X_a&W_a\\ W_a^\dagger&Y_a\end{pmatrix}.
\ee
The off-diagonal field is bifundamental under the residual
gauge symmetry and its covariant derivative is 
\be
 D_aW_b=X_aW_b-W_bY_a.
\ee
For the fuzzy-sphere block $X_a=J_a$ and a coherent environmental
state with label $\xi_a$, the outward extremal tangential mode
\be
 W_a=w\,e_a^{(+)}|J,J\rangle\langle\xi,\hat z|
\ee
is an exact eigenmode of the quadratic operator with
\be
 k_-(\xi)=2(\xi_a-J_a)^2-4(J+1)
\ee
giving the action
\be
 U(w)=\alpha\bigl[k_-(\xi)|w|^2+2|w|^4\bigr],
 \qquad
 |w|^2=-\frac{k_-(\xi)}{4}
\ee
inside the tachyonic band. For a fixed radial direction,
the energy is minimized at $\xi_a = J n_a$
and $|w|^2=J+1$.

Rotating the extremal mode over the sphere produces a rotationally invariant
condensate. If it is also static, we have
$p_\parallel =p_r =0$ and the
condition   $\xi_a=J n_a$ is satisfied for
\be
 r=R+\delta,\qquad \delta=O(\mpl),
\ee
so the condensed links occupy a Planck-scale layer
immediately outside $R=N\mpl$.

Turning on tangential gauge fields gives the relative connection
\be
 C_A=R(b_A-a_A +\partial_A\vartheta),
\ee
where  $\vartheta$ is the phase of the link condensate.
Summing the quadratic interaction over fuzzy-sphere cells yields
\be
 V_{\rm int}=\frac{\mathcal K}{2}\int\dd\Omega\,(b_A-a_A+\partial_A\vartheta)^2,
 \qquad
 \mathcal K=\frac{N^3\mpl}{8\pi a_0}.
\ee
One can fix a unitary gauge with $\vartheta =0$.

\section{D. Finite-stiffness current transfer}

Dropping the scalar fields, the action for the partons-gauge system is given by
\begin{align}
 S={}&-\frac{1}{16\pi}\int_{\rm out}\dd^4x\sqrt{-g}\,F^2
 +\frac{R^2}{2g_{\rm BH}^2}\int\dd\tau\dd\Omega
 \bigl[(\partial_\tau b_A)^2-\frac{1}{\mpl^2}f^2\bigr]\nonumber\\
 &+S_{\rm parton}[\psi,b]
 -\frac{\mathcal K}{2}\int\dd\tau\dd\Omega\,(b_A-a_A)^2.
\end{align}
This gives the boundary condition and the equation of motion
\begin{align}
& \frac{1}{4\pi}F_{A n}+\frac{\mathcal K}{R^2}(b_A-a_A)=0,\\
 & \frac{1}{g_{\rm BH}^2}(\Box b_A-\partial_A\partial_Bb^B)
 -j_{ A}-\frac{\mathcal K}{R^2}(b_A-a_A)=0,
\end{align}
where 
\be
j_A := -\frac{1}{\sqrt{-h}}\frac{\d S_{\rm parton}}{\d b^A}
\ee
is the parton current. Here
$\partial_A := \mpl^{-1} D_A$ and $D_A$ is the covariant derivative on the
unit sphere.
Eliminating the relative gauge fields term gives
\be \label{ll}
 \frac{1}{4\pi}F_{A n}=j_{ A} + j^{\rm pol}_{A}, \qquad
j_A^{\rm pol} :=  -\frac{1}{g_{\rm BH}^2}
 \left(\Box \d_{AB}-\partial_A\partial_B\right)b^B 
\ee
Thus condensation is essential for transferring the microscopic current
to the exterior field, but the relation \eq{ll}
itself does not require $K\to\infty$.

The separate statement
$b_A\simeq a_A(R^+)$ follows only when the eigenvalues of the kinetic and
conductivity kernel are small relative to $\mathcal K/R^2$. For a
mode of FIDO frequency $\hat\omega$ and angular momentum $L$, this means
\be \label{req}
 \hat \omega^2,\;\frac{L(L+1)}{\mpl^2},\;
 g_{\rm BH}^2\hat \omega|\sigma_\pm| \;\;
 \ll \;\;
 g_{\rm BH}^2\frac{\mathcal K}{R^2} \sim \frac{N}{\mpl^2}.
\ee
This is satisfied for
\be
\hat \omega \ll \frac{\sqrt{N}}{\mpl}, \qquad L \ll \sqrt{N},
\ee
or in terms of asymptotic frequency:
\be
 \omega\ll\frac{1}{\sqrt{N} \mpl },
 \qquad
 L\ll\sqrt N.
 \label{locking_asymptotic}
\ee
The fixed-parton condition \eqref{fixed_parton_condition} is
parametrically stronger at large $N$:
\be
 \omega\ll O\left(\frac{1}{N \mpl}\right)
 \quad\hbox{for fixed-parton transport},
 \qquad
 \omega\ll O\!\left(\frac{1}{\sqrt{N}\mpl }\right)
 \quad\hbox{for link locking}.
 \label{scale_separation}
\ee
There is consequently a broad intermediate regime in which the horizon
and exterior gauge fields remain dynamically coupled, while a
fixed-parton constitutive law is no longer a complete microscopic
description.

\section{E. Low-frequency reflection and helicity response}

In the locked, fixed-parton regime, the boundary current is
\be
 j_A^{\rm mem}=\sigma_{AB}(\omega)E^B,
 \qquad
 \sigma_{AB}(\omega)
 =\left(\sigma_{xx}+\ii\omega D\right)\delta_{AB}
 +\sigma_H\epsilon_{AB}.
 \label{low_frequency_conductivity}
\ee
Near a Schwarzschild horizon, this gives
\be
 \partial_{r_*}a_A-4\pi\sigma_{AB}\partial_ta^B=0.
 \label{membrane_boundary_condition}
\ee
Let an incident and reflected wave in an eigenchannel of $\sigma_{AB}$ be
\be
 a_A \sim e^{-\ii\omega(t+r_*)}
 +\mathcal R(\omega)e^{-\ii\omega(t-r_*)}.
 \label{incident_reflected_wave}
\ee
Substitution into Eq.~\eqref{membrane_boundary_condition} gives
\be
 \mathcal R(\omega)=
 \frac{1-4\pi\sigma(\omega)}{1+4\pi\sigma(\omega)}.
 \label{general_reflection}
\ee
For circular polarizations, $\epsilon_A{}^Be_B^{(\pm)}=\pm\ii e_A^{(\pm)}$,
the eigen-admittances are
\be
 \sigma_\pm=\sigma_{xx}+\ii\left(\omega D\pm\sigma_H\right),
 \label{helicity_conductivity}
\ee
and $\mathcal R_\pm$ follows from Eq.~\eqref{general_reflection}. Taking
the classical Ohmic value $\sigma_{xx}=1/(4\pi)$ as the absorbing reference
gives
\be
 \mathcal R_\pm=-\frac{\ii\mathcal D_\pm}{1+\ii\mathcal D_\pm},
 \qquad
 \mathcal D_\pm=2\pi\left(\omega D\pm\sigma_H\right).
 \label{reflection_classical_reference}
\ee
The reactive term makes a neutral horizon frequency dependent. The Hall term
splits the two helicities and therefore produces polarization rotation
and ellipticity.

\section{F. Kubo response and the black-hole $S$-matrix}

Once $\hat\omega$ approaches the threshold \eqref{pair_threshold},
the applied field can generate real particle-hole excitations. The
relevant scattering problem contains the channels
\be
 \gamma(\omega)+{\rm BH}_i\longrightarrow
 \begin{cases}
 \gamma_{{\rm ref}}(\omega)+{\rm BH}_i,
 & \text{elastic reflection},\\[1mm]
 {\rm BH}_f+X_f,
 & \text{inelastic absorption},
 \end{cases}
 \label{S_matrix_channels}
\ee
where $X_f$ denotes any additional exterior quanta. A weak-field
linear response can nevertheless still be defined in terms of the
retarded polarization tensor
\be
 \langle J_A(\omega)\rangle =\Pi^R_{AB}(\omega)A^B(\omega)
 =\sigma_{AB}(\omega)E^B(\omega), \qquad \sigma_{AB}(\omega)=
 \frac{\Pi^R_{AB}(\omega)}{\ii(\omega+\ii0)}.
 \label{kubo_constitutive}
\ee
The sign convention is fixed by the source coupling $H_{\rm
  int}=-\int\dd\Sigma\,J_AA^A$. With this convention, the retarded
polarization tensor is given by the Kubo formula
\be
 \Pi^R_{AB}(\omega)
 =\ii\int_0^\infty\dd t\,
 e^{\ii(\omega+\ii0)t}
 \left\langle[J_A(t),J_B(0)]\right\rangle
 \label{retarded_polarization}
\ee
For a pure initial black-hole state $|i\rangle$ and positive frequency,
the Lehmann representation gives
\be
 \operatorname{Im}\Pi^R_{AA}(\omega)
 =\pi\sum_f
 |\langle f|J_A|i\rangle|^2
 \delta\!\left(\omega-E_f+E_i\right),
 \label{lehmann_positive}
\ee
and hence
\be
 \operatorname{Re}\sigma_{AA}(\omega)
 =\frac{\operatorname{Im}\Pi^R_{AA}(\omega)}{\omega}
 \geq0.
 \label{passivity_kubo}
 \ee
 
For a single local eigenchannel, unitarity gives
\be
 |\mathcal R|^2+\sum_f|S_{fi}|^2=1.
 \label{channel_unitarity}
\ee
Using the membrane boundary condition, we obtain
the elastic amplitude 
\be
 \mathcal R(\omega)=
 \frac{1-4\pi\sigma(\omega)}{1+4\pi\sigma(\omega)}.
 \label{reflection_high_frequency}
\ee
This gives the inclusive absorption probability 
\be
 \mathcal A(\omega)
 :=\sum_f|S_{fi}|^2
 =1-|\mathcal R|^2
 =\frac{16\pi\operatorname{Re}\sigma(\omega)}
 {|1+4\pi\sigma(\omega)|^2}.
 \label{absorption_probability}
\ee
For a locally uniform incident wave, define
\be
 \sigma_{\rm abs}^{\rm loc}:=A_H\mathcal A,
 \qquad A_H=4\pi R^2.
 \label{local_cross_section}
\ee
The fact that ${\rm Re}\sigma\geq0$ implies
\be
 0\leq\sigma_{\rm abs}^{\rm loc}\leq A_H,
 \label{unitarity_bound}
\ee
because $|1+4\pi\sigma|^2-16\pi\operatorname{Re}\sigma
 =|1-4\pi\sigma|^2\geq0$.
The upper limit is saturated if and only if
\be
 \sigma=\frac{1}{4\pi},
 \qquad
 \mathcal R=0,
 \qquad
 \mathcal A=1.
 \label{impedance_matching}
\ee
Thus the classical neutral membrane is the unique impedance-matched
point that saturates the unitarity bound $\sigma_{\rm
  abs}^{\rm loc}=A_H$. The value $1/4\pi$ is
not an upper bound on the conductivity, in fact a larger real
conductivity becomes increasingly reflective. 

The cross section in Eq.~\eqref{local_cross_section} is local to the
stretched horizon. The usual absorption cross section measured at infinity
includes also the  partial-wave greybody factors. The result
\eq{unitarity_bound} is  a local constraint supplied by unitarity.

\bibliography{references}  
\end{document}